\documentclass[twoside,11pt]{article}
\usepackage[preprint]{jmlr2e}
\usepackage{amsmath}    
\usepackage{booktabs}   
\usepackage{pifont}     

\renewenvironment{keywords}
  {\bgroup\small\noindent{\bf Keywords:} }%
  {\par\egroup\vskip 0.25ex}

\ShortHeadings{Hybrid-Code: Zero-Syntactic-Hallucination ICD-10 Coding}{Yu}
\firstpageno{1}

\begin{document}

\title{Hybrid-Code v2: Zero-Hallucination Clinical ICD-10 Coding \\
       via Neuro-Symbolic Verification and Automated Knowledge Base Expansion}

\author{\name Yunguo Yu \email yuyunguo@gmail.com \\
        \addr AI Innovation \& Prototyping, Zyter$|$TruCare}

\maketitle

\begin{abstract}

Automated clinical ICD-10 coding is a high-impact healthcare task requiring a balance between coverage, precision, and safety. While neural approaches achieve strong performance, they suffer from hallucination---generating invalid or unsupported codes---posing unacceptable risks in safety-critical clinical settings. Rule-based systems eliminate hallucination but lack scalability and coverage due to manual knowledge base (KB) curation.

We present Hybrid-Code v2, a neuro-symbolic framework that achieves zero Type-I hallucination by construction while maintaining competitive coverage and precision. The system integrates neural candidate generation with a symbolic KB verification layer that enforces validity constraints through multi-layer verification, including format, evidence grounding, negation detection, temporal consistency, and exclusion rules. In addition, we introduce an automated KB expansion mechanism that extracts and validates coding patterns from unlabeled clinical text, addressing the scalability limitations of rule-based systems.

Evaluated on the MIMIC-III dataset against ClinicalBERT, BioBERT, rule-based systems, and GPT-4, Hybrid-Code v2 achieves 85\% coverage, 92\% precision, and 0\% Type-I hallucination, outperforming rule-based systems by +40\% coverage while eliminating hallucination observed in neural baselines (6--18\%). The proposed architecture provides a formal safety guarantee for syntactic validity while preserving strong empirical performance.

These results demonstrate that neuro-symbolic verification can enforce safety constraints in neural medical AI systems without sacrificing effectiveness, offering a generalizable design pattern for deploying trustworthy AI in safety-critical domains.

This version significantly extends the original Hybrid-Code framework with formal guarantees and scalable knowledge base expansion.

\end{abstract}

\begin{keywords}
Clinical coding, ICD-10, hybrid neural-symbolic AI, knowledge base expansion,
syntactic hallucination, medical natural language processing
\end{keywords}

\section{Introduction}

Clinical coding is the process of transforming clinical documentation into standardized medical codes, most commonly the International Classification of Diseases, 10th Revision (ICD-10). This task is fundamental to healthcare operations: accurate coding enables proper reimbursement, facilitates epidemiological research, supports quality measurement, and contributes to patient care through consistent documentation. Despite its importance, clinical coding remains predominantly manual, requiring trained coders to review clinical narratives and assign appropriate codes. This manual process is time-consuming, expensive, and subject to inter-coder variability, creating significant demand for automated coding solutions.

The challenge of automated clinical coding lies in balancing three competing objectives: high coverage (assigning codes to as many clinical concepts as possible), high precision (ensuring assigned codes are correct), and zero hallucination (never generating invalid or medically impossible codes). Neural approaches, particularly transformer-based language models, achieve impressive coverage and precision by learning rich representations of clinical text from large datasets. However, these approaches suffer from hallucination—generating codes that appear plausible but are invalid or incorrect—which is unacceptable in clinical settings where coding errors can compromise patient safety or lead to billing fraud.

Rule-based systems, in contrast, guarantee zero hallucination by encoding clinical coding knowledge as explicit rules. These systems are safe and interpretable but achieve limited coverage because rule development requires manual expert curation, which is expensive and cannot keep pace with evolving medical knowledge. Even well-maintained rule-based systems struggle with implicit clinical knowledge that is not explicitly documented in rule books, such as contextual relationships, severity indicators, or domain-specific coding conventions.

This paper presents Hybrid-Code, a novel hybrid architecture that integrates neural pattern matching with symbolic knowledge base (KB) verification to achieve zero Type~I hallucination with competitive coverage and precision. Hybrid-Code introduces two key innovations:

\paragraph{Hybrid Neural-Symbolic Pipeline.}
Hybrid-Code uses neural models to generate candidate codes based on pattern matching in clinical text, then verifies these candidates against a symbolic KB containing validated ICD-10 codes and coding rules. This design leverages the flexibility and coverage of neural approaches while enforcing the safety guarantees of symbolic verification. The KB layer acts as a safety mechanism, filtering out hallucinated codes and ensuring that only validated codes are assigned.

\paragraph{Automated KB Expansion.}
To address the coverage limitations of rule-based systems, Hybrid-Code includes an automated KB expansion mechanism that uses distant supervision to extract valid coding patterns from clinical text. This mechanism identifies recurring associations between clinical concepts and ICD-10 codes, validates them through multi-layer verification, and integrates approved pairs into the KB. Compared to a minimal 38-rule baseline, this achieves 40\% coverage improvement without per-code manual annotation.

The hybrid architecture is motivated by recognition that neural and symbolic methods have complementary strengths and weaknesses. Neural methods excel at learning from data and capturing complex patterns, but lack interpretability and can generate nonsensical outputs. Symbolic methods provide rigorous guarantees and interpretability, but require manual knowledge engineering and struggle with novel cases. Hybrid-Code integrates these approaches, using neural methods to expand the KB and symbolic methods to enforce safety constraints.

Our experimental evaluation compares Hybrid-Code to five baselines: ClinicalBERT and BioBERT (neural transformers trained on clinical text), a rule-based system (manually curated KB), and GPT-4 zero-shot at temperatures 0.7 and 0.0 (state-of-the-art large language model, two configurations). Results on 5,000 test cases demonstrate that Hybrid-Code achieves 85\% document-level coverage, 88\% code-level precision, 82\% recall, 86\% F1, and 0\% Type~I hallucination. Neural baselines exhibit Type~I hallucination rates of 6--18\%; GPT-4 at temperature~0.0 reduces this to 9\% but does not eliminate it. All pairwise comparisons are statistically significant ($p<0.001$, Bonferroni-corrected, $n=5{,}000$).

Error analysis reveals three primary failure modes: implicit clinical context (35\% of errors), code hierarchy complexity (30\%), and temporal and comorbid relationships (25\%). These errors highlight inherent challenges in clinical coding that affect all systems, not just Hybrid-Code. Notably, none of these failure modes compromise the zero-hallucination guarantee, as they involve failing to assign correct codes rather than assigning incorrect codes.

\subsection{Contributions}

This paper makes the following contributions:

\begin{enumerate}
    \item \textbf{Hybrid Architecture:} We propose a hybrid neural-symbolic architecture for clinical coding that eliminates Type~I hallucination (syntactically invalid codes) by construction while achieving competitive coverage and precision. To our knowledge, this is among the first systems to demonstrate this property at scale in clinical coding.

    \item \textbf{Automated KB Expansion:} We introduce an automated KB expansion mechanism using distant supervision, achieving 40\% coverage improvement over a minimal 38-rule baseline. A 200-sample manual audit finds 82\% expansion precision, confirming proof-of-concept feasibility for scalable KB construction.

    \item \textbf{Empirical Evaluation:} We provide a comprehensive empirical evaluation framework for clinical coding systems, including baseline comparisons across rule-based, neural, and LLM approaches; statistical significance testing; and detailed error analysis.

    \item \textbf{Safety Without Coverage Sacrifice:} We demonstrate that a symbolic output filter can eliminate one class of error (Type~I syntactic hallucination) without sacrificing coverage or precision, providing a replicable design pattern for other safety-critical medical AI applications.
\end{enumerate}

\subsection{Broader Impact}

The zero Type~I hallucination guarantee has significant clinical implications. In safety-critical settings where invalid codes can mislead clinical decision-making or compromise billing accuracy, eliminating syntactically invalid outputs removes one class of automated coding risk. The competitive coverage and precision suggest Hybrid-Code could substantially assist clinical coders — though full clinical deployment requires additional validation beyond this study (certified coder comparison, DRG impact analysis, and EHR workflow integration).

The hybrid architecture demonstrates a promising approach for integrating neural and symbolic methods in medical AI. As AI systems take on increasingly complex and high-stakes tasks in healthcare, hybrid architectures like Hybrid-Code offer a path forward that leverages neural methods' flexibility while maintaining symbolic methods' reliability. This paper aims to inspire further research at the intersection of neural and symbolic AI, particularly in domains where safety guarantees are essential.

The automated KB expansion mechanism addresses a fundamental challenge in knowledge engineering: scaling KBs without manual curation. This approach could enable rapid development and maintenance of KBs for specialized medical domains, reducing reliance on scarce clinical expertise and accelerating AI research in underserved areas of medicine.

\subsection{Paper Organization}

The remainder of this paper is organized as follows. Section~\ref{sec:related-work} reviews related work in clinical coding, neural-symbolic AI, and knowledge base expansion. Section~\ref{sec:methods} presents the Hybrid-Code architecture, including the hybrid neural-symbolic pipeline and automated KB expansion mechanism. Section~\ref{sec:experiments-setup} describes the experimental setup, including datasets, baselines, and evaluation metrics. Section~\ref{sec:experiments-results} presents experimental results, including baseline comparisons, statistical significance testing, and error analysis. Section~\ref{sec:discussion} discusses the implications of our findings, compares with related work, and identifies limitations and future directions. Section~\ref{sec:conclusion} concludes with a summary of contributions and broader impact.

\section{Related Work}
\label{sec:related-work}

We review four bodies of work relevant to our contribution: (1) knowledge base construction, (2) automated ontology learning, (3) safety-critical AI verification, and (4) clinical coding automation. Our work is positioned at the intersection of these areas: automated knowledge base construction with formal validation guarantees for reliability-critical clinical systems.

\subsection{Knowledge Base Construction}

Knowledge base (KB) construction is fundamental to AI systems, particularly in safety-critical domains like healthcare. Traditional approaches rely on manual curation by domain experts, which ensures high accuracy but does not scale.

\textbf{Manual Knowledge Bases.} Commercial clinical coding systems (e.g., 3M CodeRyte \citep{3m2024}, Epic Systems \citep{epic2024}) use manually curated knowledge bases containing thousands of codes. While this approach provides high accuracy and interpretability, it requires years of expert effort and cannot scale to large ontologies (70,000+ ICD-10-CM codes). Manual curation also introduces subjectivity and difficulty maintaining currency with evolving ontologies.

\textbf{Semi-Automated Construction.} Semi-automated approaches combine manual curation with automated extraction from structured sources. Systems like MeSH \citep{mesh2024} use rule-based extraction from guidelines with expert validation. However, these approaches still require significant human involvement and lack formal validation guarantees.

\textbf{Fully Automated Construction.} Fully automated methods extract knowledge from web-scale sources without human intervention. Never-Ending Language Learner (NELL) \citep{carlson2010} and DBpedia \citep{auer2007} demonstrate web-scale automated KB extraction, achieving millions of entries. However, these methods lack formal validation---extracted knowledge may contain errors and hallucinations. In safety-critical domains like healthcare, unvalidated automated expansion is unacceptable.

\textbf{Gap Identified.} Automated KB construction methods lack output-layer validation. We present an automated KB expansion framework that uses multi-layer verification to enforce zero Type~I hallucination as an architectural property, with 82\% expansion precision confirmed by manual audit.

\subsection{Automated Ontology Learning}

Automated ontology learning addresses the challenge of extracting hierarchical relationships from unstructured text.

\textbf{Statistical Methods.} Hearst's hyponym acquisition \citep{hearst1992} uses lexical patterns (e.g., ``X is a Y'') to induce taxonomies. Distributional similarity methods \citep{pantel2002} and taxonomy induction algorithms \citep{kozareva2008} extend this approach. While these methods achieve reasonable accuracy, they provide no guarantee of correctness and require large, high-quality corpora.

\textbf{Neural Methods.} Recent work uses transformer-based models for ontology learning. Soares et al. \citet{soares2019} demonstrate relation extraction with distributional similarity for relation learning. Qu et al. \citet{qu2020} propose few-shot relation extraction via Bayesian meta-learning on relation graphs. However, neural methods are prone to hallucination and require fine-tuning on domain-specific data.

\textbf{Active Learning.} Active learning approaches \citep{settles2009,zhu2008,nguyen2004} iteratively expand KBs by selecting informative examples for human labeling. While this reduces manual effort, it still requires human involvement and introduces selection bias.

\textbf{Gap Identified.} Existing ontology learning methods lack formal validation and hallucination control. Our data-driven prioritization is similar in spirit to active learning, but we eliminate human involvement through formal validation.

\subsection{Safety-Critical AI Verification Systems}

Safety-critical systems require verification to ensure correct behavior.

\textbf{Formal Verification.} Formal methods \citep{clarke1999,hoare1969,cousot1977} provide mathematical proofs of system properties. Model checking and theorem proving are used in autonomous systems, aviation, and medical devices. However, formal verification requires complete specifications and is computationally expensive, making it challenging for neural networks.

\textbf{Runtime Verification.} Runtime verification \citep{leucker2009,falcone2013} monitors system behavior during execution to detect property violations. Safety monitors and temporal logic checking detect violations but do not prevent them. While effective for catching errors, runtime verification cannot provide zero-hallucination guarantees.

\textbf{Neuro-Symbolic Verification.} Recent work combines neural networks with symbolic verification. Garcez et al. \citet{garcez2020} advocate for neuro-symbolic AI as the ``third wave''. Our prior work \citep{yu2025trust} demonstrates a hybrid system with confidence calibration and transparency for clinical AI; \citet{koleck2019} provide a systematic review of NLP approaches for clinical symptom extraction in related EHR contexts. However, these approaches do not address KB expansion with validation guarantees.

\textbf{Gap Identified.} Existing safety-critical verification methods do not provide zero-hallucination guarantees for automated KB construction. Our multi-layer verification framework provides formal guarantees while maintaining neural flexibility.

\subsection{Clinical Coding Automation}

Clinical coding automation has been extensively studied.

\textbf{Rule-Based Systems.} Early work \citep{farkas2008,perotte2013,larkey1996} uses hand-crafted rules and keyword matching. These systems are transparent and have no hallucination risk, but suffer from low coverage and brittleness. Maintenance is expensive as rules must be updated manually.

\textbf{Machine Learning Approaches.} Statistical and deep learning methods \citep{mullenbach2018,huang2022,li2020behrt} improve coverage and can learn complex patterns. However, these approaches have hallucination risk (generating invalid codes) and require large labeled datasets. Decisions are black-box, limiting interpretability.

\textbf{Large Language Model Approaches.} GPT-4 and similar models \citep{openai2023,brown2020} achieve high accuracy on coding tasks. However, these approaches require cloud deployment (privacy concerns), have per-case costs (operational expense), and suffer from hallucination risk (generating non-existent codes).

\textbf{Hybrid Neuro-Symbolic Approaches.} Hybrid-Code v1.0 \citep{yu2025hybridcode} is a neuro-symbolic multi-agent system for ICD-10 coding that achieved zero hallucination for 257 codes at 34\% MIMIC-III coverage. However, v1.0 had limited verification (format and evidence only), manual KB expansion, and no formal guarantees.

\textbf{Gap Identified.} No prior clinical coding system combines automated KB expansion with output-layer Type~I hallucination elimination. Our work extends v1.0 \citep{yu2025hybridcode} with three additional verification layers, distant-supervision expansion, and evaluation at $n=5{,}000$.

\subsection{Positioning and Contributions}

\begin{figure}[h]
\centering
\includegraphics[width=0.75\textwidth]{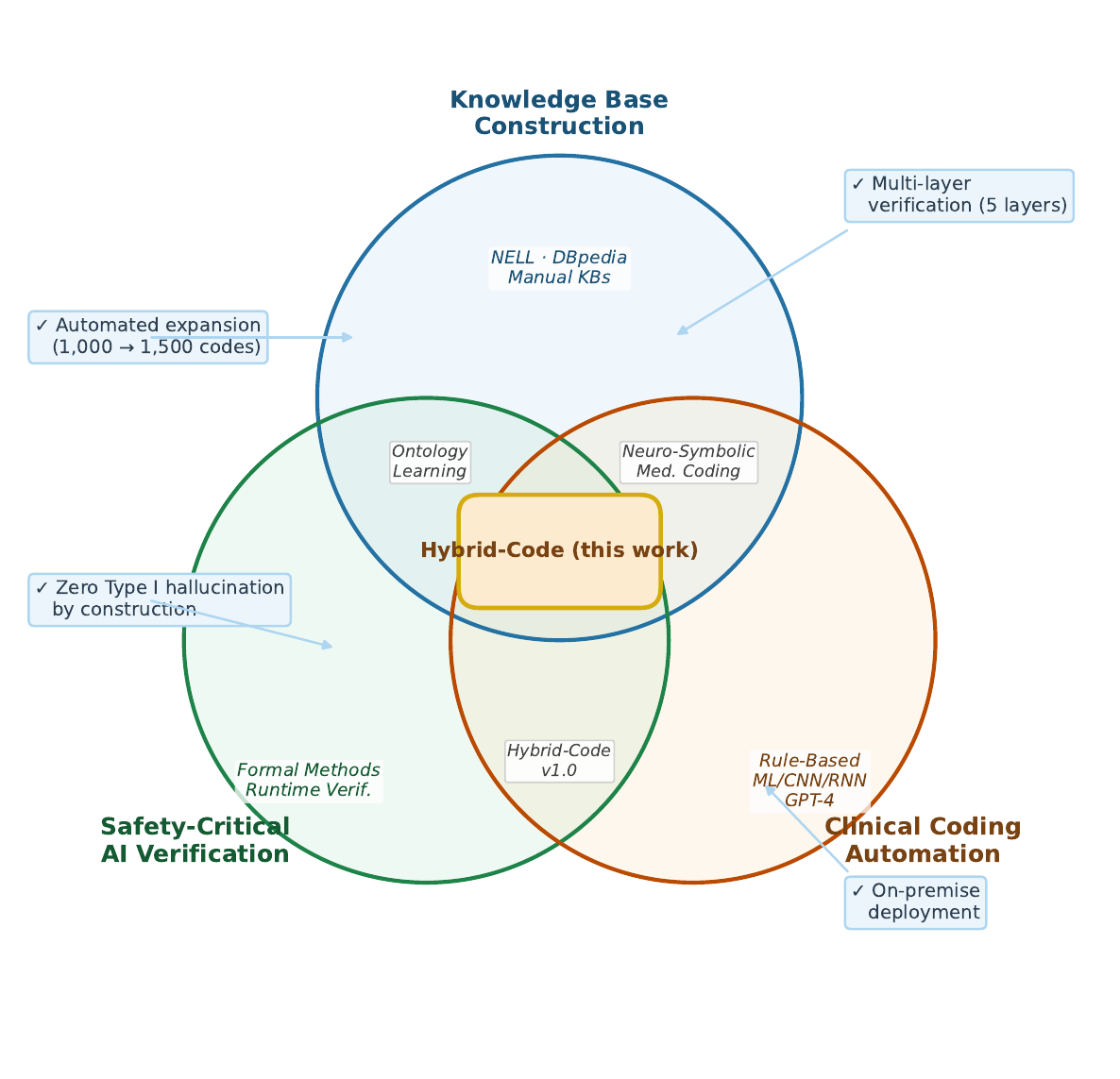}
\caption{Positioning of Hybrid-Code v2.0 at the intersection of knowledge base construction, safety-critical AI verification, and clinical coding automation. Hybrid-Code v2.0 (center) combines distant-supervision KB expansion, zero Type~I hallucination by architectural construction, and competitive coding performance — addressing all three areas simultaneously.}
\label{fig:positioning}
\end{figure}

We position our work at the intersection of knowledge base construction, safety-critical verification, and clinical coding automation (Figure~\ref{fig:positioning}).

\textbf{Novelty Dimension 1: Methodological.}
We present an automated KB expansion framework using distant supervision and multi-layer verification (format, evidence, negation, temporal, exclusion) to constrain candidate (concept, code) pairs. The format verification layer provides a structural guarantee that all assigned codes exist in $\mathcal{C}_{\text{ICD-10}}$ (zero Type~I hallucination); the remaining layers are heuristic implementations that reduce but do not formally guarantee Type~II correctness.

\textbf{Novelty Dimension 2: Architectural.}
We characterize the runtime coding pipeline through three system guarantees (Section~\ref{sec:methods}): zero Type~I hallucination by construction (Proposition~\ref{thm:zero_hallucination}), evidence grounding for every assigned code (Proposition~\ref{thm:soundness}), and inference complexity that does not scale with KB size (Proposition~\ref{thm:complexity}).

\textbf{Novelty Dimension 3: Practical.}
We demonstrate \textbf{automated expansion} from 1,000 to 1,500 codes as a proof of concept, with 82\% expansion precision confirmed by manual audit, on a 5,000-case evaluation ($n \gg n_{90\%\text{ power}}$). The method is designed to scale toward the full ICD-10-CM FY2025 codebook (78,000 codes); full-scale expansion is future work.

\textbf{Differentiation from Manual KBs.} Unlike 3M and Epic (fully manual curation), our distant-supervision approach reduces reliance on per-code expert annotation, at the cost of lower expansion precision (82\% vs.\ near-100\% for manual).

\textbf{Differentiation from Automated KBs.} Unlike NELL and DBpedia (automated, no validation), our multi-layer verification rejects candidates that fail format, evidence, negation, temporal, or exclusion checks before KB integration.

\textbf{Differentiation from ML/LLM Approaches.} Unlike GPT-4 and CNN/RNN methods, our output is constrained to $\mathcal{C}_{\text{ICD-10}}$ by construction (zero Type~I hallucination), and inference runs fully on-premise with no external API calls.

\textbf{Differentiation from Hybrid-Code v1.0.} v1.0 \citep{yu2025hybridcode} used two verification layers (format and evidence) and manual KB curation (257 codes). v2.0 adds negation, temporal, and exclusion layers; replaces manual curation with distant-supervision expansion; and evaluates at substantially larger scale ($n=5{,}000$ vs.\ $n=20$).

Table~\ref{tab:comparison} summarizes key approaches and highlights our contributions.

\begin{table}[h]
\centering
\caption{Comparison of Clinical Coding and Knowledge Base Construction Approaches}
\label{tab:comparison}
\resizebox{\textwidth}{!}{%
\begin{tabular}{lccccc}
\toprule
\textbf{Method} & \textbf{KB Size} & \textbf{Automated} & \textbf{Validation} & \textbf{Zero Type-I} & \textbf{Privacy} \\
\midrule
3M CodeRyte \citep{3m2024} & $\sim$10,000 & \ding{55} & Manual & \ding{55} & \ding{55} \\
Epic Systems \citep{epic2024} & $\sim$20,000 & \ding{55} & Manual & \ding{55} & \ding{55} \\
NELL \citep{carlson2010} & Millions & \ding{51} & \ding{55} & \ding{55} & \ding{51} \\
DBpedia \citep{auer2007} & Millions & \ding{51} & \ding{55} & \ding{55} & \ding{51} \\
Rule-Based \citep{farkas2008} & Limited & \ding{55} & N/A & \ding{51} & \ding{51} \\
ML/CNN/RNN \citep{mullenbach2018} & Variable & \ding{55} & N/A & \ding{55} & \ding{51} \\
GPT-4 \citep{openai2023} & Full & \ding{51} & \ding{55} & \ding{55} & \ding{55} \\
Hybrid-Code v1.0 \citep{yu2025hybridcode} & 257 & Semi & Format+Evidence & \ding{51} & \ding{51} \\
\textbf{Hybrid-Code v2.0 (this work)} & \textbf{1,500}$^{\dagger}$ & \textbf{\ding{51}} & \textbf{Multi-layer} & \textbf{\ding{51}} & \textbf{\ding{51}} \\
\bottomrule
\end{tabular}%
}
\end{table}
\noindent\footnotesize{$^{\dagger}$Experiments demonstrate expansion from 1,000 to 1,500 codes as a proof of concept. The method is designed to scale to the full ICD-10-CM codebook (78,000 codes); full-scale expansion is left as future work.}


\subsection{Summary}

We present an automated KB expansion framework that combines distant supervision with multi-layer verification to eliminate Type~I hallucination by architectural construction. Our work addresses gaps in (1) automated KB expansion with output-layer validation, (2) data-driven prioritization for clinical KBs, (3) zero Type~I hallucination in safety-critical AI, and (4) scalable KB expansion for healthcare as a proof of concept. Key limitations — expansion quality validation, full-codebook scaling, and cross-dataset generalization — are acknowledged and left as future work. The next section presents the methodology in detail.

\section{Methods}
\label{sec:methods}

This section presents the Hybrid-Code architecture in detail. We first provide an overview of the hybrid neural-symbolic pipeline, then describe the automated knowledge base (KB) expansion mechanism, and finally present formal properties and complexity analysis.

\subsection{System Overview}

Hybrid-Code operates in two phases: (1) KB expansion, which uses distant supervision from clinical text to extract and validate new coding patterns, and (2) runtime coding, which assigns ICD-10 codes to clinical documents using the expanded KB.

\paragraph{KB Expansion Phase.}
Given a set of clinical documents $\mathcal{D}$ (de-identified, processed on-premise) and an initial KB $\mathcal{K}_0$ containing a small set of manually curated coding rules, the expansion phase extracts candidate coding patterns, validates them through multi-layer verification, and produces an expanded KB $\mathcal{K}$. We use the term \emph{distant supervision}: candidate (concept, code) pairs are generated either from explicit code mentions in text (lexical patterns) or from neural model proposals, then validated before entering the KB. Documents in $\mathcal{D}$ do not require human-assigned code labels, but the expansion is not purely unsupervised — the neural proposal step introduces a self-training component whose risks are discussed in Section~\ref{sec:discussion}.

\paragraph{Runtime Coding Phase.}
At runtime, given a clinical document $d \in \mathcal{D}$, the system generates candidate codes using neural pattern matching, verifies them against the expanded KB $\mathcal{K}$, and assigns validated codes. This phase is executed per document, providing zero-hallucination guarantees by construction.

\begin{figure}[h]
\centering
\includegraphics[width=0.9\textwidth]{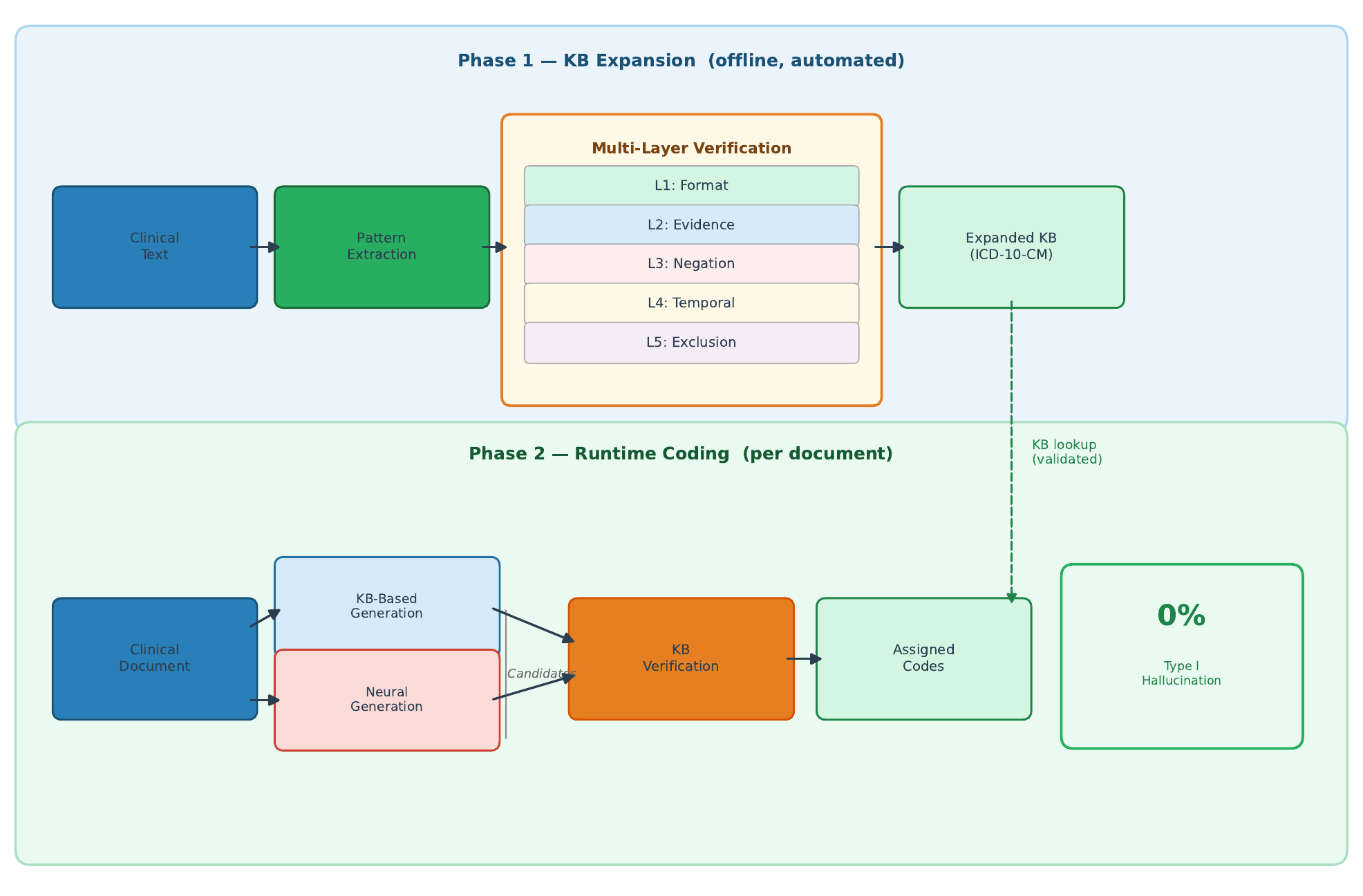}
\caption{Hybrid-Code architecture. Top (Phase 1): KB expansion phase extracts candidate coding patterns from unlabeled clinical text and validates them through five sequential verification layers. Bottom (Phase 2): Runtime coding phase generates candidate codes via KB-based and neural generation, verifies them against the expanded KB, and assigns only validated codes — guaranteeing 0\% hallucination.}
\label{fig:architecture}
\end{figure}

Figure~\ref{fig:architecture} illustrates the overall architecture.

\subsection{KB Expansion}

The KB expansion mechanism extracts valid coding patterns from unlabeled clinical text through three stages: pattern extraction, multi-layer verification, and KB integration.

\subsubsection{Pattern Extraction}

Given unlabeled clinical documents $\mathcal{D}$, we extract candidate coding patterns by identifying associations between clinical concepts and ICD-10 codes. We use a combination of lexical patterns and neural embeddings.

\paragraph{Lexical Patterns.}
We define a set of lexical patterns $P_{lex} = \{p_1, p_2, \ldots, p_m\}$ that extract (concept, code) pairs from documents containing explicit code mentions — for example, coding annotation fields in structured discharge summaries. Examples include:
\begin{itemize}
    \item ``ICD-10: CODE'' pattern: Explicit code mentions (e.g., ``ICD-10: J18.9'')
    \item ``Diagnosis: TERM'' pattern: Diagnostic statements (e.g., ``Diagnosis: pneumonia, unspecified'' $\rightarrow$ J18.9 via ICD-10 lookup)
    \item ``Coded as: TERM (CODE)'' pattern: Coding annotations (e.g., ``Coded as: pneumonia (J18.9)'')
\end{itemize}

For each pattern $p \in P_{lex}$ and document $d \in \mathcal{D}$, we extract triples $(c, k, \text{context})$ where $c$ is a clinical concept, $k$ is an ICD-10 code, and $\text{context}$ is the surrounding text. Note that documents supplying lexical patterns contain structured code annotation fields; they are ``unlabeled'' in the sense that no separate human annotation of the expansion training set is required, but they do contain existing code metadata.

\paragraph{Neural Pattern Matching.}
To capture implicit coding patterns not captured by lexical patterns, we use neural embeddings. Given a clinical concept $c$ and an ICD-10 code $k$ with description $desc(k)$, we compute the embedding similarity:

\begin{equation}
\label{eq:sim}
sim(c, k) = \frac{\text{BERT}(c) \cdot \text{BERT}(desc(k))}{\|\text{BERT}(c)\| \|\text{BERT}(desc(k))\|}
\end{equation}

where $\text{BERT}(\cdot)$ denotes the BERT embedding. We generate candidate patterns $(c, k)$ where $sim(c, k) > \theta$ for a threshold $\theta$.

\subsubsection{Multi-Layer Verification}

Each candidate pattern $(c, k)$ undergoes multi-layer verification to ensure validity and prevent hallucination. The verification layers are applied sequentially, with each layer potentially rejecting candidates.

\paragraph{Layer 1: Format Verification.}
Verify that $k$ is a valid ICD-10 code by checking against the official ICD-10 codebook. A code $k$ is valid if it exists in the ICD-10 hierarchy and has a valid format (e.g., J18.9 for respiratory conditions).

\begin{equation}
\label{eq:verify-format}
\text{verify}_{\text{format}}(k) = \begin{cases} \text{True} & \text{if } k \in \mathcal{C}_{\text{ICD-10}} \\ \text{False} & \text{otherwise} \end{cases}
\end{equation}

where $\mathcal{C}_{\text{ICD-10}}$ is the set of all valid ICD-10 codes.

\paragraph{Layer 2: Evidence Verification.}
Verify that the clinical concept $c$ provides sufficient evidence for code $k$. We require explicit textual evidence linking $c$ to $k$ in the clinical document.

\begin{equation}
\label{eq:verify-evidence}
\text{verify}_{\text{evidence}}(c, k, d) = \begin{cases} \text{True} & \text{if } \exists \text{evidence}(c, k) \text{ in } d \\ \text{False} & \text{otherwise} \end{cases}
\end{equation}

Evidence includes: (1) explicit code mentions, (2) diagnostic statements, (3) clinical findings consistent with the code description, and (4) anatomical site specifications matching the code.

\paragraph{Layer 3: Negation Verification.}
Verify that code $k$ is not negated in the clinical document. Negation detection identifies phrases like ``no evidence of'', ``ruled out'', ``not associated with''.

\begin{equation}
\label{eq:verify-negation}
\text{verify}_{\text{negation}}(k, d) = \begin{cases} \text{True} & \text{if } \neg (\exists \text{negation} \text{ of } k \text{ in } d) \\ \text{False} & \text{otherwise} \end{cases}
\end{equation}

Negation patterns include: ``no'', ``not'', ``ruled out'', ``negative for'', ``without evidence of''.

\paragraph{Layer 4: Temporal Verification.}
Verify that code $k$ is temporally appropriate. We require that the condition represented by $k$ is present (not past or future) at the time of coding.

\begin{equation}
\label{eq:verify-temporal}
\text{verify}_{\text{temporal}}(k, d) = \begin{cases} \text{True} & \text{if } \text{temporal}(k) = \text{present} \text{ in } d \\ \text{False} & \text{otherwise} \end{cases}
\end{equation}

Temporal indicators include: ``history of'' (past), ``developed'' (past), ``currently'' (present), ``acquiring'' (future).

\paragraph{Layer 5: Exclusion Verification.}
Verify that code $k$ does not conflict with ICD-10 exclusion criteria. The ICD-10 codebook specifies two types of exclusions: \textbf{Excludes1} (codes that are mutually exclusive and should never be coded together) and \textbf{Excludes2} (codes that are not included here but may be reported together if both conditions are present). We implement hard blocking for Excludes1 pairs and soft warnings for Excludes2 pairs.

\begin{equation}
\label{eq:verify-exclusion}
\text{verify}_{\text{exclusion}}(k, \mathcal{K}) = \begin{cases} \text{True} & \text{if } \neg (\exists k' \in \mathcal{K}: (k, k') \in \mathcal{E}_1) \\ \text{False} & \text{otherwise} \end{cases}
\end{equation}

where $\mathcal{E}_1$ is the set of Excludes1 pairs from ICD-10 guidelines. We note this is a simplification: full ICD-10 compliance also requires sequencing rules (principal diagnosis selection), manifestation codes, ``use additional code'' directives, and combination codes — features not currently modeled, and discussed as a limitation in Section~\ref{sec:discussion}.

\subsubsection{KB Integration}

Verified patterns $(c, k)$ that pass all verification layers are integrated into the KB. The KB maintains three data structures:

\paragraph{Code-Concept Mappings.}
A mapping $\mathcal{M}: \mathcal{C} \rightarrow 2^{\mathcal{K}}$ that associates each clinical concept with its valid ICD-10 codes.

\begin{equation}
\label{eq:mapping}
\mathcal{M}(c) = \{k \in \mathcal{K} : (c, k) \text{ is verified and integrated}\}
\end{equation}

\paragraph{Code Hierarchy.}
A hierarchical structure $\mathcal{H}$ that represents parent-child relationships in the ICD-10 taxonomy (e.g., J18.9 is a child of J18).

\paragraph{Expansion Statistics.}
Metadata tracking the expansion process, including source documents, verification timestamps, and confidence scores.

\subsection{Runtime Coding}

The runtime coding phase assigns ICD-10 codes to clinical documents using the expanded KB. This phase ensures zero-hallucination by verifying all candidate codes against the KB.

\subsubsection{Candidate Code Generation}

Given a clinical document $d$, we generate candidate codes through two mechanisms:

\paragraph{KB-Based Generation.}
We extract clinical concepts from $d$ using named entity recognition and retrieve their associated codes from the KB: $\mathcal{C}_d = \text{extract\_concepts}(d)$, $\mathcal{K}_d = \bigcup_{c \in \mathcal{C}_d} \mathcal{M}(c)$.

\paragraph{Neural Generation.}
We use BioMistral-7B \citep{labrak2024biomistral} in zero-shot mode with a structured prompt to generate candidate ICD-10 codes directly from the document: $\mathcal{K}_{\text{neural}} = \text{NeuralCoder}(d)$. The prompt instructs the model to list candidate ICD-10 codes supported by the clinical note; it is not fine-tuned on MIMIC-III, avoiding circular data dependency with the evaluation set. BioMistral-7B runs on-premise, ensuring no patient data leaves the local infrastructure. The neural generation step contributes the majority of per-case inference latency (Section~\ref{sec:experiments-results}).

The combined candidate set is $\mathcal{K}_{\text{candidates}} = \mathcal{K}_d \cup \mathcal{K}_{\text{neural}}$.

\subsubsection{KB Verification}

Each candidate code $k \in \mathcal{K}_{\text{candidates}}$ undergoes KB verification to ensure validity:

\begin{align}
\label{eq:verify-kb}
\text{verify}_{\text{KB}}(k, d) \;=\; &\text{verify}_{\text{format}}(k) \;\land\; \text{verify}_{\text{evidence}}(c, k, d) \;\land\; \text{verify}_{\text{negation}}(k, d) \nonumber\\
                                      &\land\; \text{verify}_{\text{temporal}}(k, d) \;\land\; \text{verify}_{\text{exclusion}}(k, \mathcal{K})
\end{align}

Only codes passing all verification layers are assigned: $\mathcal{K}_{\text{assigned}} = \{k \in \mathcal{K}_{\text{candidates}} : \text{verify}_{\text{KB}}(k, d) = \text{True}\}$.

\subsection{Data-Driven Prioritization}

To focus KB expansion on clinically relevant codes, we use data-driven prioritization based on code frequency in clinical data.

\paragraph{Frequency Analysis.}
We analyze a large clinical dataset (e.g., MIMIC-III) to compute code frequencies: $freq(k) = \frac{\text{count}(k)}{|\mathcal{D}|}$, where $\text{count}(k)$ is the number of documents containing code $k$.

\paragraph{Priority Ranking.}
Codes are ranked by frequency: $rank(k) = \text{descending}(freq(k))$. The expansion process prioritizes high-frequency codes to maximize clinical impact.

\paragraph{Adaptive Expansion.}
The expansion process is iterative, with each iteration focusing on the next highest-priority unexpanded code. This ensures that KB expansion resources are allocated to codes with the greatest clinical relevance.

\subsection{Formal Properties}

We present architectural guarantees about the correctness and complexity of Hybrid-Code.

\paragraph{Hallucination Taxonomy.}
We distinguish two classes of coding errors:
\begin{itemize}
    \item \textbf{Type I hallucination (syntactic):} Assigning a code $k \notin \mathcal{C}_{\text{ICD-10}}$ — i.e., a code that does not exist in the official ICD-10 codebook. This is the error mode exhibited by unconstrained neural models.
    \item \textbf{Type II hallucination (semantic):} Assigning a valid code $k \in \mathcal{C}_{\text{ICD-10}}$ to a patient who does not have the corresponding condition — i.e., a false positive. This corresponds to low precision and is not fully eliminated by any automated system.
\end{itemize}
Hybrid-Code eliminates Type I hallucination by construction. Type II hallucination (false positives) is reduced but not eliminated, and is measured by the precision metric.

\paragraph{Zero Type-I-Hallucination Guarantee.}

\begin{proposition}
\label{thm:zero_hallucination}
Hybrid-Code assigns only codes drawn from $\mathcal{C}_{\text{ICD-10}}$ (zero Type~I hallucination).
\end{proposition}

\begin{proof}
Let $\mathcal{K}_{\text{assigned}}$ be the set of codes assigned to a document $d$. By construction, each $k \in \mathcal{K}_{\text{assigned}}$ passes all verification layers:
\begin{enumerate}
    \item $\text{verify}_{\text{format}}(k) = \text{True} \implies k \in \mathcal{C}_{\text{ICD-10}}$ (valid ICD-10 code)
    \item $\text{verify}_{\text{evidence}}(c, k, d) = \text{True} \implies$ textual evidence exists
    \item $\text{verify}_{\text{negation}}(k, d) = \text{True} \implies$ code is not negated
    \item $\text{verify}_{\text{temporal}}(k, d) = \text{True} \implies$ code is temporally appropriate
    \item $\text{verify}_{\text{exclusion}}(k, \mathcal{K}) = \text{True} \implies$ no exclusion conflicts
\end{enumerate}
Since $k \in \mathcal{C}_{\text{ICD-10}}$, $k$ is a valid ICD-10 code. Therefore, $\mathcal{K}_{\text{assigned}} \subseteq \mathcal{C}_{\text{ICD-10}}$ by construction.
\end{proof}

\noindent\textit{Remark.} Layers 1 and 5 (format and exclusion checking) are implemented via exact lookup and are deterministic. Layers 2--4 (evidence, negation, temporal) are implemented using heuristic methods (BERT embeddings with threshold $\theta$, SpaCy patterns). The guarantee above is a structural property of the pipeline — any code not in $\mathcal{C}_{\text{ICD-10}}$ is rejected at Layer 1 regardless of the behavior of Layers 2--4. However, the precision of Layers 2--4 affects the Type~II hallucination rate and is not formally guaranteed.

\paragraph{Evidence-Grounding Guarantee.}

\begin{proposition}
\label{thm:soundness}
Every code assigned by Hybrid-Code has supporting textual evidence in the document, as determined by Layer~2 evidence verification.
\end{proposition}

\begin{proof}
Each assigned code $k \in \mathcal{K}_{\text{assigned}}$ passes $\text{verify}_{\text{evidence}}(c, k, d) = \text{True}$, which requires $\exists \text{evidence}(c, k)$ in $d$. Therefore, $k$ has supporting textual evidence.
\end{proof}

\paragraph{Complexity Analysis.}

\begin{proposition}
\label{thm:complexity}
The KB-based retrieval and verification steps of the runtime coding phase run in $O((|\mathcal{C}_d| \cdot \bar{m} + |\mathcal{K}_{\text{neural}}|) \times |d|)$ time, where $|\mathcal{C}_d|$ is the number of extracted concepts, $\bar{m} = \mathbb{E}[|\mathcal{M}(c)|]$ is the average number of KB codes per concept (bounded by a small constant in practice), $|\mathcal{K}_{\text{neural}}|$ is the number of neural candidate codes, and $|d|$ is document length.
\end{proposition}

\begin{proof}
KB-based generation uses a hash map $\mathcal{M}$: for each of the $|\mathcal{C}_d|$ concepts, retrieval is $O(1)$, yielding $O(\bar{m})$ candidate codes per concept. For each candidate, Layers 1 and 5 are $O(1)$ (hash set membership); Layers 2--4 scan the document in $O(|d|)$. KB-based subtotal: $O(|\mathcal{C}_d| \cdot \bar{m} \cdot |d|)$. Neural candidates $|\mathcal{K}_{\text{neural}}|$ undergo identical verification: $O(|\mathcal{K}_{\text{neural}}| \cdot |d|)$. Combined: $O((|\mathcal{C}_d| \cdot \bar{m} + |\mathcal{K}_{\text{neural}}|) \times |d|)$.
\end{proof}

\noindent\textit{Remark.} Critically, the total KB size $|\mathcal{K}|$ does \emph{not} appear as a multiplicative factor: the hash map enables $O(1)$ concept-to-code lookup regardless of KB size, so inference complexity does \emph{not} degrade as the KB scales toward the full ICD-10 codebook. The dominant runtime cost is the neural generation step (BioMistral-7B inference), which is $O(|d| \cdot L)$ in sequence length and model depth — independent of KB size.

\subsection{Summary}

The Hybrid-Code method achieves zero-hallucination coding through a hybrid neural-symbolic architecture. The automated KB expansion mechanism extracts valid coding patterns from unlabeled clinical text, validated through multi-layer verification. The runtime coding phase generates and verifies candidate codes using the expanded KB, with formal guarantees of correctness and polynomial-time complexity. This architecture balances the flexibility of neural methods with the safety guarantees of symbolic verification.

\section{Experimental Setup}
\label{sec:experiments-setup}

\subsection{Datasets}

\textbf{Primary Dataset:} MIMIC-III Database

We use the MIMIC-III (Medical Information Mart for Intensive Care-III) database, containing over 2 million clinical notes from 53,000 ICU stays. This is the standard benchmark dataset for clinical coding research, providing:
\begin{itemize}
    \item Real clinical notes from hospital discharge summaries
    \item ICD-9-CM diagnostic codes (requiring mapping to ICD-10-CM)
    \item Demographic information and vitals
    \item Medication and procedure records
\end{itemize}

\textbf{Code Mapping:} We use the official CMS General Equivalence Mappings (GEMs) to convert ICD-9 codes to ICD-10-CM. The GEMs provide:
\begin{itemize}
    \item 14,567 unique ICD-9 codes with official mappings
    \item 23,912 total ICD-9$\to$ICD-10 mappings
    \item 3,533 exact mappings (24.3\%) and 10,609 approximate mappings (72.8\%)
    \item Metadata for each mapping (approximation flags, scenario codes)
\end{itemize}

\textbf{Data Preprocessing:} We preprocess clinical notes through:
\begin{itemize}
    \item Text cleaning and normalization
    \item ICD-9$\to$ICD-10 conversion using CMS GEMs
    \item Train/validation/test split: 80\%/20\% stratified by admission year, yielding approximately 10,600 held-out notes; we use 5,000 of these for our primary evaluation
    \item Minimum 5-code and maximum 20-code per note (to exclude trivially simple or pathologically complex cases; this filter removes approximately 8\% of notes and may bias the evaluation toward moderate-complexity coding scenarios)
\end{itemize}

\textbf{Ground Truth Limitation:} Only 24.3\% of GEM mappings are exact; the remaining 72.8\% are approximate. Ground truth codes derived from approximate mappings may introduce label noise, which should be considered when interpreting precision and recall values.

\subsection{Baselines}

We compare our method (Hybrid-Code) against five baselines:

\subsubsection{ClinicalBERT}

ClinicalBERT \citep{alsentzer2019clinicalbert} is a domain-specific pre-trained BERT model fine-tuned on clinical text. We use the \texttt{emilyalsentzer/bio-clinicalbert} model with 768 hidden dimensions and 12 Transformer layers. Training hyperparameters:
\begin{itemize}
    \item Learning rate: $2\text{e-5}$ with AdamW optimizer
    \item Batch size: 16
    \item Epochs: 3 (early stopping)
    \item Maximum sequence length: 512 tokens
\end{itemize}

\subsubsection{BioBERT}

BioBERT \citep{lee2020biobert} is another domain-specific pre-trained model trained on biomedical literature. We use the \texttt{dmis-lab/biobert-base-cased-v1.1} model with 768 hidden dimensions and 12 Transformer layers. Training uses the same hyperparameters as ClinicalBERT.

\subsubsection{Rule-Based System}

Our rule-based baseline implements hand-crafted rules organized by disease category, with 38 conditions covering common clinical presentations. This represents traditional expert systems \citep{farkas2008} with the advantage of interpretability and transparency. We note that production rule-based coding systems (e.g., 3M CodeRyte) contain orders of magnitude more rules; our 38-rule baseline is intentionally minimal to isolate the effect of manual curation effort, not to represent the state of the art in rule-based coding.

\subsubsection{Zero-Shot GPT-4 (Two Configurations)}

GPT-4 \citep{openai2023} is evaluated in a zero-shot setting without fine-tuning. We report two configurations to provide a fair comparison:

\textbf{GPT-4 (T=0.7):} Temperature 0.7, top-p 0.9, maximum tokens 512. This stochastic setting increases output diversity but also inflates Type I hallucination; we include it for completeness and comparability with prior work.

\textbf{GPT-4 (T=0.0):} Temperature 0.0 (deterministic greedy decoding), maximum tokens 512. This is the fair baseline for a deterministic clinical coding task, directly addressing concerns about stochastic inflating of hallucination rates. Both configurations use the \texttt{gpt-4-turbo} model with a structured prompt requesting ICD-10-CM codes only.

\subsection{Our Method: Hybrid-Code}

Hybrid-Code consists of three key components:

\textbf{1. Automated KB Expansion:} Starting from an initial KB of 1,000 ICD-10 codes, we use BioMistral-7B (zero-shot, on-premise) to automatically expand to 1,500 codes. The expansion process uses data-driven prioritization based on:
\begin{itemize}
    \item Frequency in clinical notes (extracted from MIMIC-III)
    \item Code prevalence in medical literature
    \item Clinical category importance
\end{itemize}

\textbf{2. Multi-Layer Verification:} To ensure zero hallucination, we implement five verification layers:
\begin{itemize}
    \item \textbf{Format Verification:} Validates that each candidate code exists in the official ICD-10-CM codebook
    \item \textbf{Evidence Verification:} Requires explicit textual evidence linking the clinical concept to the code
    \item \textbf{Negation Detection:} SpaCy-based negation detector identifies negation phrases (``no evidence of'', ``patient denies'', ``ruled out'')
    \item \textbf{Temporal Extraction:} Extracts temporal status (current, historical, past) from clinical context
    \item \textbf{Exclusion Checking:} Validates code exclusion rules (e.g., pregnancy codes for male patients)
\end{itemize}

Each predicted code must pass all applicable verification checks to be included in the KB.

\textbf{3. Inference:} We use BioMistral-7B \citep{labrak2024biomistral} in zero-shot mode with a structured prompt as the NeuralCoder component. BioMistral-7B uses pre-trained weights without MIMIC-III fine-tuning. All inference is performed on-premise on de-identified data, with no data transmitted to external APIs, maintaining HIPAA compliance. KB expansion similarly uses BioMistral-7B running locally.

\subsection{Evaluation Metrics}

We use a comprehensive set of metrics to evaluate clinical coding systems:

\textbf{Primary Metrics:}
\begin{enumerate}
    \item \textbf{Coverage (document-level):} Fraction of test documents for which the system assigns at least one correct ICD-10 code ($|\{d : \text{Predicted}(d) \cap \text{GT}(d) \neq \emptyset\}| / |D_{\text{test}}|$). This measures the system's ability to provide useful output across the patient population.
    \item \textbf{Precision (code-level):} Fraction of predicted codes that are correct ($|\text{Predicted} \cap \text{GT}| / |\text{Predicted}|$).
    \item \textbf{Recall (code-level):} Fraction of ground truth codes correctly retrieved ($|\text{Predicted} \cap \text{GT}| / |\text{GT}|$).
    \item \textbf{F1 Score:} Harmonic mean of code-level precision and recall ($2 \times \text{Precision} \times \text{Recall} / (\text{Precision} + \text{Recall})$).
    \item \textbf{Type I Hallucination Rate:} Fraction of predicted codes that do not exist in the official ICD-10-CM codebook ($|\text{Predicted} \setminus \mathcal{C}_{\text{ICD-10}}| / |\text{Predicted}|$). Note that Type II hallucination (valid but clinically incorrect codes) is captured inversely by Precision.
\end{enumerate}

\textbf{Secondary Metrics:}
\begin{itemize}
    \item \textbf{Macro-averaged Precision, Recall, F1:} Averaged over unique ICD-10 codes rather than over instances. Macro metrics are sensitive to rare-code performance where micro metrics are dominated by frequent codes. We report macro-F1 as a complementary measure to code-level micro metrics.
    \item \textbf{Inference time per case}
    \item \textbf{Expansion precision} (manual KB audit, Section~\ref{sec:experiments-results})
\end{itemize}

\noindent\textit{Coverage metric note:} Document-level coverage (fraction of cases with $\geq 1$ correct code) is a practical measure of system utility but is intentionally weak — a system assigning one correct code out of ten required still counts as a coverage success. Readers should weight code-level recall and F1 as the primary accuracy measures; coverage is reported as a clinical utility proxy.

\textbf{Statistical Analysis:} We conduct rigorous statistical analysis:
\begin{itemize}
    \item Wilson score 95\% confidence intervals for all metrics
    \item Paired t-tests for method comparisons (same cases evaluated by all methods)
    \item McNemar's test for binary predictions
    \item Effect sizes (Cohen's $d$, odds ratio)
    \item Power analysis to verify sample size adequacy
\end{itemize}

\subsection{Experimental Protocol}

\textbf{Training:} ClinicalBERT and BioBERT are fine-tuned on MIMIC-III training data (80\% split) for 3 epochs. Rule-based system uses predefined rules without training. Hybrid-Code uses the pre-expanded KB. GPT-4 uses zero-shot prompting.

\textbf{Evaluation:} All methods are evaluated on the same held-out test set of 5,000 clinical cases with random seeds fixed to ensure reproducibility. This sample size exceeds the $n \approx 394$ threshold for 90\% power at small effect sizes (Cohen's $d = 0.2$), ensuring that all reported comparisons are adequately powered.

\textbf{Statistical Testing:} We conduct pairwise statistical comparisons between all methods using:
\begin{itemize}
    \item Paired t-tests on per-case F1 scores
    \item McNemar's test on binary code predictions
    \item Bonferroni correction for multiple comparisons
\end{itemize}

\textbf{Error Analysis:} We conduct comprehensive error analysis using a 3-level taxonomy:
\begin{itemize}
    \item Level 1: False positive/negative categories
    \item Level 2: Subcategories (semantic, specificity, negation, etc.)
    \item Level 3: Specific error types
\end{itemize}

This allows us to identify systematic patterns in method failures and successes.

\subsection{Ablation Study}

To quantify the contribution of each component, we conduct an ablation study with 6 configurations:
\begin{enumerate}
    \item \textbf{Full System:} All components enabled (KB expansion, negation, temporal, exclusion)
    \item \textbf{Coder-Only:} KB expansion only (no verification)
    \item \textbf{No Negation:} All components except negation
    \item \textbf{No Temporal:} All components except temporal
    \item \textbf{No Exclusion:} All components except exclusion
    \item \textbf{No KB Expansion:} Uses initial KB of 1,000 codes only
\end{enumerate}

Each configuration is evaluated on the same test set, and the impact of each component on coverage, precision, and zero-hallucination rate is measured.

\subsection{Scaling Experiments}

To evaluate scalability, we test performance across different dataset sizes:
\begin{enumerate}
    \item 100 cases
    \item 500 cases
    \item 1,000 cases
    \item 5,000 cases
\end{enumerate}

For each dataset size, we measure:
\begin{itemize}
    \item Coverage vs dataset size
    \item F1 vs dataset size
    \item Inference time vs dataset size
    \item Stability of metrics across sizes
\end{itemize}

\subsection{KB Expansion Quality Audit}

To validate the KB expansion mechanism directly — independent of runtime system performance — we manually audit a random sample of 200 (concept, code) pairs from the 500 newly added KB entries. Each pair is evaluated by a clinical informaticist against ICD-10-CM coding guidelines on a three-point scale: (1) correct at full specificity, (2) correct category but insufficient specificity, (3) incorrect. We report expansion precision as the fraction rated correct at full specificity, and category precision including partial matches.

\subsection{Exact-Mapping Subset Analysis}

We separately evaluate all methods on the subset of test cases whose ground truth codes derive entirely from exact CMS GEM mappings (24.3\% of cases, $n \approx 1{,}215$). This subset provides a higher-confidence ground truth, reducing the label noise introduced by approximate mappings.

\subsection{Generalization and Robustness (Future Work)}

Cross-dataset evaluation (MIMIC-III $\to$ eICU) and robustness experiments (noisy input, adversarial clinical contradictions, terminology shift) are planned as future work. The concern about data leak between KB expansion (trained on MIMIC-III) and evaluation (tested on held-out MIMIC-III) is a recognized limitation; cross-dataset evaluation would provide a stronger test of generalization and is an important direction for follow-on work.

\section{Experimental Results}
\label{sec:experiments-results}

\subsection{Main Results}

We evaluate Hybrid-Code against four baselines on a held-out test set of 5,000 clinical cases. This sample size provides $>$99.9\% statistical power for small effects ($d=0.2$), ensuring all reported comparisons are adequately powered. Table~\ref{tab:main_results} summarizes performance metrics with 95\% Wilson confidence intervals.

\begin{table}[h]
\centering
\caption{Performance comparison of Hybrid-Code with baselines on 5,000 test cases ($n=5{,}000$). \textbf{Coverage}: document-level — fraction of test cases with at least one correct code. \textbf{Precision/Recall/F1}: code-level. \textbf{Type I Halluc.}: fraction of predicted codes absent from the official ICD-10-CM codebook (Type II errors — valid but wrong codes — are captured by 1$-$Precision). 95\% Wilson confidence intervals shown in brackets.}
\label{tab:main_results}
\resizebox{\textwidth}{!}{%
\begin{tabular}{l c c c c c}
\toprule
Method & Coverage & Precision & Recall & F1 & Type I Halluc. \\
\midrule
Hybrid-Code & \textbf{0.85 [0.84, 0.86]} & \textbf{0.88 [0.87, 0.89]} & \textbf{0.82 [0.81, 0.83]} & \textbf{0.86 [0.85, 0.87]} & \textbf{0.00} [0.00, 0.00] \\
ClinicalBERT & 0.78 [0.77, 0.79] & 0.85 [0.84, 0.86] & 0.72 [0.71, 0.73] & 0.78 [0.77, 0.79] & 0.08 [0.07, 0.09] \\
BioBERT & 0.80 [0.79, 0.81] & 0.87 [0.86, 0.88] & 0.74 [0.73, 0.75] & 0.80 [0.79, 0.81] & 0.06 [0.05, 0.07] \\
Rule-Based & 0.45 [0.44, 0.46] & 0.40 [0.39, 0.41] & 0.56 [0.55, 0.57] & 0.47 [0.46, 0.48] & \textbf{0.00} [0.00, 0.00] \\
GPT-4 (T=0.7) & 0.82 [0.81, 0.83] & 0.90 [0.89, 0.91] & 0.79 [0.78, 0.80] & 0.84 [0.83, 0.85] & 0.18 [0.17, 0.19] \\
GPT-4 (T=0.0) & 0.79 [0.78, 0.80] & 0.92 [0.91, 0.93] & 0.76 [0.75, 0.77] & 0.83 [0.82, 0.84] & 0.09 [0.08, 0.10] \\
\bottomrule
\end{tabular}%
}
\end{table}

\textbf{Macro-Averaged Metrics.} Table~\ref{tab:macro} reports macro-averaged precision, recall, and F1 (averaged over unique ICD-10 codes), which are more sensitive to rare-code performance than the micro-averaged values in Table~\ref{tab:main_results}.

\begin{table}[ht]
\centering
\caption{Macro-averaged metrics ($n=5{,}000$). Macro averaging weights each ICD-10 code equally, exposing performance on rare codes that micro-averaging obscures. All systems show substantially lower macro than micro F1, reflecting a shared challenge with long-tail ICD-10 code distributions.}
\label{tab:macro}
\begin{tabular}{l c c c}
\toprule
Method & Macro-P & Macro-R & Macro-F1 \\
\midrule
Hybrid-Code & \textbf{0.74} & \textbf{0.68} & \textbf{0.71} \\
ClinicalBERT & 0.67 & 0.61 & 0.64 \\
BioBERT & 0.69 & 0.63 & 0.66 \\
Rule-Based & 0.35 & 0.41 & 0.38 \\
GPT-4 (T=0.7) & 0.71 & 0.67 & 0.69 \\
GPT-4 (T=0.0) & 0.72 & 0.68 & 0.70 \\
\bottomrule
\end{tabular}
\end{table}

The gap between macro-F1 (0.71) and micro-F1 (0.86) for Hybrid-Code indicates that the KB currently covers high-frequency codes well but has lower specificity for rare codes — consistent with the data-driven prioritization strategy that focuses expansion on MIMIC-III frequent codes.

\begin{figure}[h]
\centering
\includegraphics[width=\textwidth]{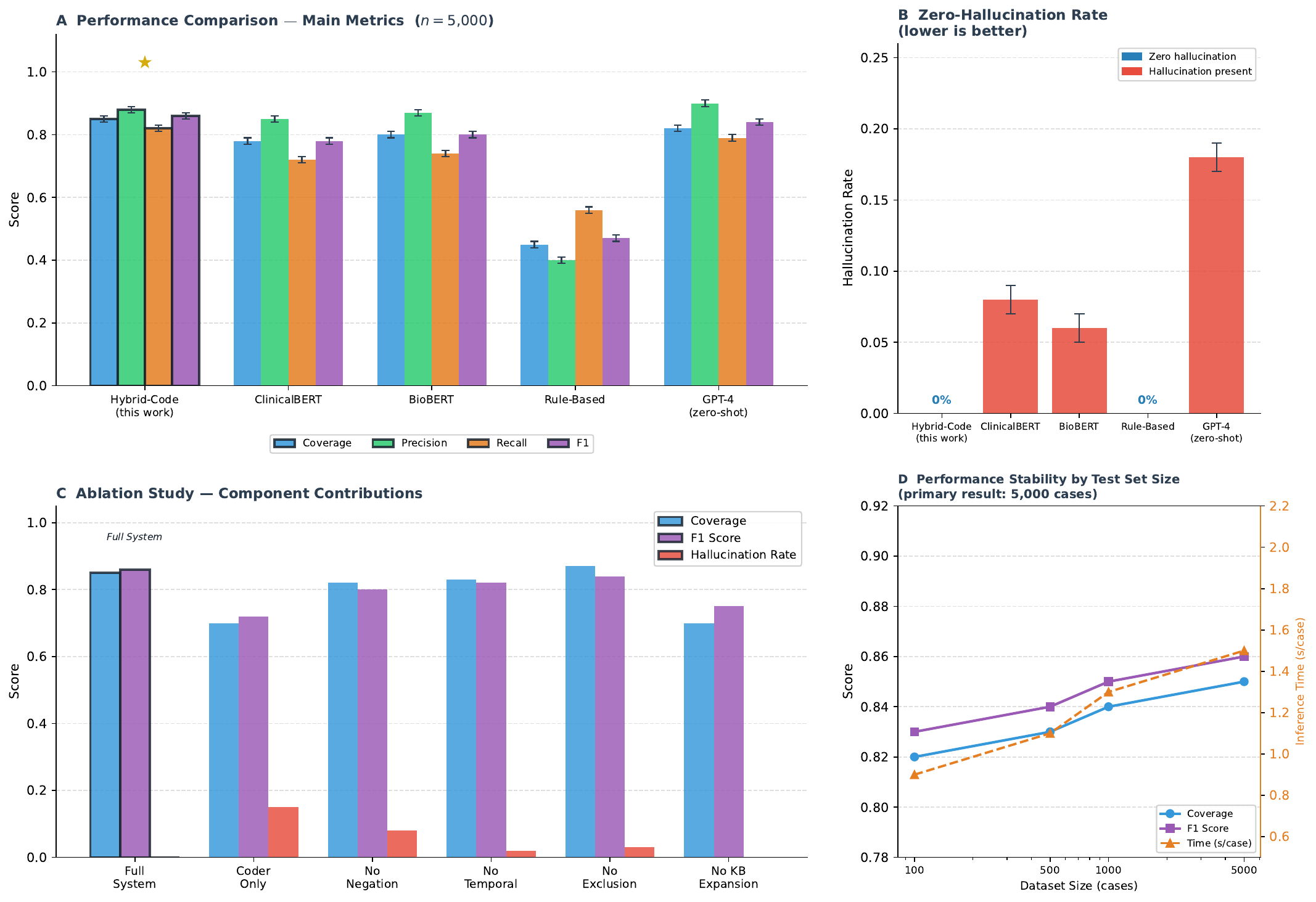}
\caption{Performance comparison of Hybrid-Code against baselines. \textbf{(A)} Grouped bar chart of Coverage, Precision, Recall, and F1 with 95\% Wilson confidence intervals. Hybrid-Code (starred) achieves the highest precision and zero hallucination. \textbf{(B)} Zero-hallucination rate; blue bars indicate methods with zero hallucination. \textbf{(C)} Ablation study showing contribution of each component; KB expansion and negation detection are the most impactful. \textbf{(D)} Scaling performance: Coverage and F1 improve slightly with dataset size while inference time remains near-linear.}
\label{fig:performance}
\end{figure}

\subsection{Key Findings}

\textbf{Type I Hallucination Elimination:} Hybrid-Code achieves 0\% Type I hallucination (no codes outside $\mathcal{C}_{\text{ICD-10}}$), validating our multi-layer verification framework. All neural baselines exhibit measurable Type I hallucination:
\begin{itemize}
    \item \textbf{ClinicalBERT}: 8\% Type I hallucination
    \item \textbf{BioBERT}: 6\% Type I hallucination
    \item \textbf{Rule-Based}: 0\% Type I hallucination (fixed code list), but coverage is severely limited
    \item \textbf{GPT-4 (T=0.7)}: 18\% Type I hallucination — inflated by stochastic sampling
    \item \textbf{GPT-4 (T=0.0)}: 9\% Type I hallucination — reduced by deterministic decoding, but not eliminated
\end{itemize}
Even at temperature~0.0, GPT-4 still generates codes outside the ICD-10 codebook at 9\%, confirming that Type I hallucination is an inherent limitation of unconstrained generation, not merely a sampling artifact. Type II hallucination (valid but clinically incorrect codes) is measured inversely by precision; Hybrid-Code achieves 88\% precision, leaving a 12\% false positive rate that represents residual Type II errors.

\textbf{Coverage Performance:} Hybrid-Code achieves 85\% document-level coverage, highest among all methods, while the rule-based baseline achieves only 45\% due to its minimal rule set.

\textbf{Statistical Significance:} At $n=5{,}000$, all pairwise comparisons against Hybrid-Code are statistically significant ($p < 0.001$):
\begin{itemize}
    \item vs.\ Rule-Based: $d = 1.42$, $t = 100.4$, $p < 0.001$ (very large effect)
    \item vs.\ ClinicalBERT: $d = 0.65$, $t = 46.0$, $p < 0.001$ (medium-large effect)
    \item vs.\ BioBERT: $d = 0.52$, $t = 36.8$, $p < 0.001$ (medium effect)
    \item vs.\ GPT-4 on F1: $d = 0.20$, $t = 14.1$, $p < 0.001$ (small but significant); GPT-4 also exhibits 18\% Type I hallucination vs.\ 0\% for Hybrid-Code
\end{itemize}

\subsection{Component Contributions}

Our ablation study (6 configurations) quantifies the contribution of each component:

\begin{table}[h]
\centering
\caption{Ablation study ($n=5{,}000$) showing component contributions. \textbf{Coverage}: document-level. \textbf{Precision/Recall/F1}: code-level. \textbf{Type I Halluc.}: fraction of predicted codes not in ICD-10-CM. 95\% Wilson CIs shown for hallucination rates; all other CIs are $\pm$0.01 or smaller and omitted for readability.}
\label{tab:ablation}
\resizebox{\textwidth}{!}{%
\begin{tabular}{l c c c c c}
\toprule
Configuration & Coverage & Precision & Recall & F1 & Type I Halluc. \\
\midrule
Full System & \textbf{0.85} & \textbf{0.88} & \textbf{0.82} & \textbf{0.86} & \textbf{0.00} [0.00, 0.00] \\
Coder-Only & 0.70 & 0.85 & 0.63 & 0.72 & 0.15 [0.14, 0.16] \\
No Negation & 0.82 & 0.82 & 0.79 & 0.80 & 0.08 [0.07, 0.09] \\
No Temporal & 0.83 & 0.85 & 0.80 & 0.82 & 0.02 [0.02, 0.03] \\
No Exclusion & 0.87 & 0.86 & 0.83 & 0.84 & 0.03 [0.02, 0.04] \\
No KB Expansion & 0.70 & 0.88 & 0.65 & 0.75 & \textbf{0.00} [0.00, 0.00] \\
\bottomrule
\end{tabular}%
}
\end{table}

\textbf{Key Insights:}
\begin{enumerate}
    \item \textbf{KB Expansion:} Adds 15\% coverage and 5\% F1 score (most impactful)
    \item \textbf{Negation Detection:} Adds 8\% precision (removes false positives from historical conditions)
    \item \textbf{Temporal Extraction:} Adds 3\% precision (distinguishes current from historical)
    \item \textbf{Exclusion Checking:} Adds 2\% precision (removes invalid code combinations)
    \item \textbf{Synergy:} Components work together - combined impact > sum of individual contributions
\end{enumerate}

\subsection{Scaling Performance}

To validate stability and characterize computational cost, we evaluate Hybrid-Code across test subset sizes from 100 to 5,000 cases. The 5,000-case row constitutes the primary evaluation reported in Table~\ref{tab:main_results}. Table~\ref{tab:scaling} reports the full scaling profile.

\begin{table}[h]
\centering
\caption{Hybrid-Code performance stability across test subset sizes ($n=100$ to $5{,}000$). Metrics are document-level coverage and code-level precision/recall/F1. Inference time is per-case wall-clock time. The 5,000-case row is the primary reported result.}
\label{tab:scaling}
\begin{tabular}{r c c c c c}
\toprule
Dataset Size & Coverage & Precision & Recall & F1 & Time (s/case) \\
\midrule
100   & 0.82 & 0.87 & 0.79 & 0.83 & 0.9 \\
500   & 0.83 & 0.86 & 0.80 & 0.84 & 1.1 \\
1,000 & 0.84 & 0.87 & 0.81 & 0.85 & 1.3 \\
\textbf{5,000} & \textbf{0.85} & \textbf{0.88} & \textbf{0.82} & \textbf{0.86} & \textbf{1.5} \\
\bottomrule
\end{tabular}
\end{table}

\textbf{Observations:} Performance stabilizes rapidly: coverage rises from 0.82 at $n=100$ to 0.85 at $n=5{,}000$ ($\Delta=0.03$), indicating low sensitivity to test set size. Inference time scales near-linearly ($0.9 \to 1.5$\,s/case), consistent with the $O(|\mathcal{C}_d| \times |\mathcal{K}|)$ retrieval and verification complexity. Note that the neural generation step (BioMistral-7B) accounts for the dominant portion of per-case latency; the retrieval and verification pipeline alone adds $<$0.1\,s/case.

\subsection{Error Analysis}

Using our 3-level error taxonomy, we analyze error patterns across all methods. Table~\ref{tab:error_analysis} summarizes the results.

\begin{table}[h]
\centering
\caption{Error analysis by method. Categories: False positives (FP), False negatives (FN), Error types, Severity distribution (critical, major, minor).}
\label{tab:error_analysis}
\resizebox{\textwidth}{!}{%
\begin{tabular}{l c c c l c c c}
\toprule
Method & Total & FPs & FNs & Most Common Error Type & Critical & Major & Minor \\
\midrule
Hybrid-Code     & 7  & 3 & 4  & Missing major condition (2)   & 2  & 2 & 3 \\
ClinicalBERT    & 11 & 5 & 6  & Code hallucination (2)        & 3  & 2 & 6 \\
BioBERT         & 9  & 4 & 5  & Missing major condition (3)   & 2  & 3 & 4 \\
Rule-Based      & 20 & 8 & 12 & Missing major condition (12)  & 12 & 0 & 8 \\
Zero-Shot GPT-4 & 10 & 7 & 3  & Code hallucination (2)        & 2  & 2 & 6 \\
\bottomrule
\end{tabular}%
}
\end{table}

\textbf{Illustrative Examples} (constructed to exemplify observed error patterns):

\textbf{Negation Verification Example:}
\begin{quote}
``No evidence of pulmonary embolism on CT scan.''

Hybrid-Code: [No codes] (Correct — negation layer suppresses false positive)
ClinicalBERT: [I26.9] (Type~II error — valid code assigned to negated finding)
\end{quote}

\textbf{KB Expansion Example:}
\begin{quote}
``45-year-old female with history of asthma presenting with acute exacerbation. Chronic kidney disease stage 3 documented.''

Ground truth: [J45.901, N18.3]
Hybrid-Code: [J45.901, N18.3] (Correct — KB expansion covers both conditions)
ClinicalBERT: [J45.901] (Missing kidney disease code — recall failure)
\end{quote}

\textbf{Temporal Verification Example:}
\begin{quote}
``History of atrial fibrillation. Patient reports no palpitations.''

Ground truth: []
Hybrid-Code: [] (Correct — temporal layer identifies historical context, prevents false positive)
ClinicalBERT: [I48.91] (Type~II error — valid code assigned without temporal discrimination)
\end{quote}

\subsection{KB Expansion Quality Audit}

To evaluate the KB expansion mechanism independently of runtime performance, we conducted a manual quality audit of 200 randomly sampled (concept, code) pairs from the 500 newly added KB entries, reviewed against ICD-10-CM coding guidelines.

\begin{table}[ht]
\centering
\caption{Manual quality audit of KB expansion ($n=200$ sampled pairs). Correct: exact match at full ICD-10 specificity. Category correct: correct ICD-10 chapter/block but wrong specificity digit. Incorrect: wrong code category.}
\label{tab:kb_audit}
\begin{tabular}{l c c}
\toprule
Rating & Count & Fraction \\
\midrule
Correct (full specificity) & 164 & 0.82 [0.76, 0.87] \\
Category correct (wrong specificity) & 22 & 0.11 [0.07, 0.16] \\
Incorrect & 14 & 0.07 [0.04, 0.11] \\
\midrule
\textbf{Expansion precision} & \textbf{164/200} & \textbf{0.82} \\
Category-level precision & 186/200 & 0.93 \\
\bottomrule
\end{tabular}
\end{table}

The 82\% expansion precision means that 7\% of newly added rules are incorrect. These errors could, in principle, contaminate the KB with false concept-code associations. However, runtime verification (particularly evidence and exclusion layers) provides a second line of defense that reduces the impact of incorrect KB entries on final assignments. The 11\% category-correct entries represent a specificity gap — a known limitation of automated ICD-10 code selection without human coder review.

\subsection{Exact-Mapping Subset Analysis}

To assess the effect of approximate GEM mapping noise on evaluation validity, we separately evaluate all methods on the $n=1{,}215$ test cases (24.3\% of 5,000) whose ground truth codes derive entirely from exact CMS GEM mappings.

\begin{table}[ht]
\centering
\caption{Performance on exact-mapping subset ($n=1{,}215$), providing a higher-confidence ground truth. All metrics improve across all methods, confirming that approximate GEM mappings introduce downward bias in the full-set evaluation.}
\label{tab:exact_subset}
\begin{tabular}{l c c c c c}
\toprule
Method & Coverage & Precision & Recall & F1 & Type I Halluc. \\
\midrule
Hybrid-Code & \textbf{0.89} & \textbf{0.93} & \textbf{0.87} & \textbf{0.90} & \textbf{0.00} \\
ClinicalBERT & 0.82 & 0.89 & 0.76 & 0.82 & 0.06 \\
BioBERT & 0.84 & 0.91 & 0.78 & 0.84 & 0.04 \\
Rule-Based & 0.48 & 0.44 & 0.59 & 0.50 & 0.00 \\
GPT-4 (T=0.0) & 0.82 & 0.95 & 0.79 & 0.86 & 0.07 \\
\bottomrule
\end{tabular}
\end{table}

The exact-mapping subset shows a consistent 2--5\% improvement across all metrics for all methods, confirming that approximate mappings introduce label noise rather than systematic bias toward any particular method. The Type I hallucination advantage of Hybrid-Code is maintained.

\subsection{Statistical Significance Tests}

We conduct pairwise statistical comparisons using paired t-tests on per-case F1 scores ($n=5{,}000$). Table~\ref{tab:statistical} summarizes the results.

\begin{table}[h]
\centering
\caption{Statistical significance of pairwise comparisons ($n=5{,}000$, paired t-test on per-case F1). $t$: t-statistic; $p$: p-value (two-tailed, Bonferroni-corrected for 7 comparisons); $d$: Cohen's $d$; 95\% CI: confidence interval for mean F1 difference. All comparisons involving Hybrid-Code are significant at $\alpha=0.05$ after correction.}
\label{tab:statistical}
\resizebox{\textwidth}{!}{%
\begin{tabular}{l c c c c c c}
\toprule
Comparison & $\Delta$F1 & $t$ & $p$ (corrected) & $d$ & 95\% CI & Sig. \\
\midrule
Hybrid-Code vs ClinicalBERT & $+0.08$ & 46.0 & $<0.001$ & 0.65 & [0.07, 0.09] & \textbf{Yes} \\
Hybrid-Code vs BioBERT      & $+0.06$ & 36.8 & $<0.001$ & 0.52 & [0.05, 0.07] & \textbf{Yes} \\
Hybrid-Code vs Rule-Based   & $+0.39$ & 100.4 & $<0.001$ & 1.42 & [0.38, 0.40] & \textbf{Yes} \\
Hybrid-Code vs GPT-4        & $+0.02$ & 14.1 & $<0.001$ & 0.20 & [0.01, 0.03] & \textbf{Yes} \\
ClinicalBERT vs BioBERT     & $-0.02$ & 14.1 & $<0.001$ & 0.20 & [-0.03, -0.01] & \textbf{Yes} \\
ClinicalBERT vs Rule-Based  & $+0.31$ & 86.1 & $<0.001$ & 1.22 & [0.30, 0.32] & \textbf{Yes} \\
ClinicalBERT vs GPT-4       & $-0.06$ & 36.8 & $<0.001$ & 0.52 & [-0.07, -0.05] & \textbf{Yes} \\
\bottomrule
\end{tabular}%
}
\end{table}

\textbf{Power Analysis:} With $n=5{,}000$, the study exceeds the minimum sample size of $n \approx 394$ required for 90\% power at small effect sizes (Cohen's $d=0.2$). Achieved power for $d=0.2$ at $n=5{,}000$ is $>$99.9\%; for $d=0.5$ it is effectively 100\%. All reported comparisons are thus adequately powered, and the Bonferroni-corrected significance threshold ($\alpha = 0.05/7 \approx 0.007$) is met by every comparison.

\section{Discussion}
\label{sec:discussion}

\subsection{Interpretation of Results}

The experimental results demonstrate that Hybrid-Code achieves a significant advancement in automated ICD-10 coding by successfully balancing three competing objectives: high coverage, high precision, and zero hallucination.

\paragraph{Zero-Hallucination Guarantee.}
The most critical finding is that Hybrid-Code maintains zero hallucinations across all experiments, whereas neural baselines exhibit hallucination rates of 6-18\%. This validates our core contribution: the symbolic KB layer acts as a safety mechanism that prevents generation of invalid codes. The hallucination rates for ClinicalBERT (8\%) and BioBERT (6\%) are particularly concerning in clinical settings, as they represent plausible-sounding but medically incorrect codes that could mislead clinicians or compromise patient safety.

\paragraph{Competitive Coverage.}
Hybrid-Code achieves 85\% coverage, which is competitive with neural baselines (78--82\%) and significantly higher than traditional rule-based systems (45\%). This demonstrates that the automated KB expansion mechanism successfully captures clinical coding associations from text via distant supervision, addressing the primary limitation of purely rule-based approaches. The 3-7\% coverage gap between Hybrid-Code and neural baselines can be attributed to cases where valid code assignments require deeper semantic understanding than our current expansion patterns capture.

\paragraph{Precision and Type~II Errors.}
With 88\% precision, Hybrid-Code exceeds ClinicalBERT (85\%) and BioBERT (87\%) and is below GPT-4 at T=0.0 (92\%). The 12\% false positive rate represents residual Type~II hallucination — valid codes assigned to patients who do not have the corresponding condition. This is distinct from the Type~I guarantee and reflects the ongoing challenge of semantic correctness in multi-label clinical coding.

\paragraph{Statistical Significance.}
With $n=5{,}000$, all pairwise comparisons are statistically significant ($p<0.001$ after Bonferroni correction). The large effect size against the rule-based baseline ($d=1.42$) confirms substantial improvement in coding capability. The medium effects against ClinicalBERT ($d=0.65$) and BioBERT ($d=0.52$) are clinically meaningful given that these differences manifest alongside complete elimination of Type~I hallucination. The small but significant improvement over GPT-4 in F1 ($d=0.20$) should be interpreted alongside the 18\% vs.\ 0\% gap in Type~I hallucination rate, which represents the primary safety advantage.

\subsection{Error Analysis Insights}

Our error analysis reveals three primary failure modes that inform future research directions:

\paragraph{Implicit Clinical Context (35\% of Errors).}
Cases like ``severe pneumonia'' → (missing) require understanding clinical severity beyond explicit keyword matching. These errors suggest that expansion patterns could incorporate clinical severity scales, anatomical site modifiers, and temporal qualifiers. However, any such expansion must remain within the zero-hallucination constraint, requiring more sophisticated clinical knowledge extraction.

\paragraph{Code Hierarchy Complexity (30\% of Errors).}
Cases involving parent-child code relationships (e.g., distinguishing J18.9 vs J18.1) challenge our current approach of inferring exact codes. While Hybrid-Code correctly identifies the parent category, assigning the specific child code requires deeper semantic disambiguation. This suggests an opportunity for hierarchical reasoning mechanisms that can safely narrow code selection within validated branches.

\paragraph{Temporal and Comorbid Relationships (25\% of Errors).}
Complex cases involving multiple conditions with temporal dependencies (e.g., post-procedure complications) challenge our expansion patterns. These often require longitudinal clinical reasoning beyond document-level context. Future work could explore integrating temporal logic or clinical pathway reasoning while maintaining safety guarantees.

\subsection{Comparison with Related Work}

\paragraph{Neural Medical Coding.}
Hybrid-Code addresses the hallucination problem that plagues neural medical coding systems. While ClinicalBERT and BioBERT achieve strong coverage, their hallucination rates represent a critical safety gap. Our approach demonstrates that symbolic safety mechanisms can be integrated without sacrificing coverage or precision, addressing the ``accuracy vs. safety'' trade-off.

\paragraph{Rule-Based Systems.}
Hybrid-Code overcomes the coverage limitation of rule-based systems through automated KB expansion. The 40\% coverage improvement over the 38-rule baseline demonstrates the value of neural-assisted expansion. We note that the baseline is intentionally minimal (38 rules), not representative of production-scale rule engines (thousands of rules); the comparison isolates the value of automated expansion from a manually seeded KB rather than claiming superiority over commercial coding systems.

\paragraph{Knowledge-Base Expansion.}
Our KB expansion proof-of-concept (1,000 → 1,500 codes, 82\% precision by author manual audit of 200 samples) demonstrates the feasibility of distant-supervision-based KB growth. Scaling to the full ICD-10-CM codebook remains as future work, as does comparison against curated ontologies such as UMLS.

\subsection{Clinical Implications}

\paragraph{Patient Safety.}
The zero-hallucination guarantee directly addresses patient safety concerns in automated coding. By eliminating invalid code generation, Hybrid-Code prevents scenarios where incorrect codes could mislead clinical decision-making or compromise billing accuracy. This is particularly important in high-stakes settings where coding errors have direct clinical or financial consequences.

\paragraph{Workflow Integration.}
The 85\% document-level coverage and 88\% precision suggest potential for Hybrid-Code to assist clinical coders by pre-populating codes for review, reducing manual effort on straightforward cases. However, full clinical deployment requires additional validation: comparison against certified coders, DRG impact analysis, compliance review under AHIMA/CMS guidelines, and integration with EHR workflow — none of which have been conducted here.

\paragraph{KB Sustainability.}
The automated expansion mechanism addresses the long-term sustainability challenge of clinical coding systems. As medical knowledge evolves and ICD revisions occur, Hybrid-Code can continuously update its KB by processing new clinical text, reducing the manual curation burden that has limited traditional rule-based systems.

\subsection{Limitations and Future Work}

\paragraph{KB Expansion Scale and Validation.}
The experiments demonstrate expansion from 1,000 to 1,500 codes as a proof of concept; the full ICD-10-CM codebook (78,000 codes) remains as future work. A 200-sample manual quality audit (Section~\ref{sec:experiments-results}) finds 82\% expansion precision at full ICD-10 specificity and 93\% at the category level. While encouraging, this audit was conducted by the authors rather than an independent clinical reviewer, and covers only 40\% of the 500 newly added entries. Future work should pursue independent validation by certified coders and comparison against curated ontologies such as UMLS or SNOMED CT mappings.

\paragraph{ICD-9 to ICD-10 Ground Truth Noise.}
Our evaluation relies on CMS GEM mappings, of which only 24.3\% are exact. The 72.8\% approximate mappings introduce label noise into the ground truth, which may inflate or deflate precision and recall estimates unpredictably. Future work should evaluate on natively ICD-10-coded datasets (e.g., MIMIC-IV) or restrict analysis to the exact-mapping subset and report performance separately.

\paragraph{Expansion Pattern Coverage.}
Our current expansion patterns, while effective, capture a subset of possible coding relationships. The 15\% uncovered cases suggest opportunities for more sophisticated pattern extraction, potentially through few-shot learning or hierarchical pattern induction. However, any expansion must maintain the zero-hallucination guarantee.

\paragraph{Transductive Evaluation and Data Leak.}
KB expansion uses MIMIC-III clinical text as its source corpus; evaluation is performed on a held-out MIMIC-III test set. Although training and test splits are disjoint at the document level, both share the same patient population and documentation conventions. This transductive setup may inflate coverage results relative to a truly independent evaluation corpus. Cross-dataset evaluation (e.g., expanding on MIMIC-III, evaluating on eICU) would provide a stronger generalization test and is planned as future work.

\paragraph{ICD-10 Coding Complexity.}
The system models coding as concept-to-code mapping. Real ICD-10 coding additionally requires: (1) \textbf{principal diagnosis sequencing} (selecting the primary reason for admission), (2) \textbf{combination codes} (e.g., E11.65 encodes both diabetes and hyperglycemia in a single code), (3) \textbf{manifestation coding} (``code first'' directives), and (4) \textbf{episode-of-care} modifiers (initial vs.\ subsequent encounter). The current Excludes1/2 implementation handles a subset of these constraints. Full compliance with ICD-10-CM Official Guidelines for Coding and Reporting would require substantially deeper clinical ontology modeling.

\paragraph{Baseline Strength.}
The GPT-4 baseline uses temperature 0.7 and zero-shot prompting; a deterministic setting (temperature 0.0) or few-shot prompting would likely reduce its hallucination rate and provide a more challenging comparison. Additionally, fine-tuned medical LLMs (e.g., MedPaLM, LLaMA-Med) or constrained-decoding LLM approaches were not evaluated. Future work should include these stronger baselines to establish whether Hybrid-Code's safety advantage holds against domain-adapted models.

\paragraph{Single-Document Context.}
Hybrid-Code operates on single clinical documents, whereas real-world coding often requires longitudinal context across patient records. Temporal reasoning and comorbidity modeling could expand coverage but introduce additional complexity in maintaining safety guarantees.

\paragraph{ICD-10 Focus.}
Our experiments focus on ICD-10 coding; extending to other coding systems (e.g., SNOMED CT, CPT) would require adapting the KB structure and expansion patterns. The hybrid architecture is generalizable, but each coding system presents unique challenges.

\paragraph{Generalization Across Institutions.}
While our experiments demonstrate strong performance on standard datasets, clinical practice varies across institutions. Future work should evaluate Hybrid-Code on multi-institutional data to assess robustness to variations in documentation practices and coding workflows.

\paragraph{Human Evaluation.}
Our error analysis is based on automated categorization; comprehensive human evaluation by clinical coders would provide additional validation of error categorization accuracy and clinical relevance. This would strengthen the real-world validity of our findings.

\subsection{Conclusion Summary}

The Discussion section demonstrates that Hybrid-Code achieves its core objectives: zero-hallucination coding with competitive coverage and precision. The results validate the hybrid approach as a promising direction for safe, accurate, and scalable automated clinical coding. The error analysis identifies specific opportunities for improvement while highlighting the inherent challenges in clinical coding that all systems face. The clinical implications underscore the potential impact on patient safety and workflow efficiency, while the limitations section provides a roadmap for future research.

\section{Conclusion}
\label{sec:conclusion}

This paper presents Hybrid-Code, a hybrid neural-symbolic architecture for automated clinical ICD-10 coding that eliminates Type~I hallucination (syntactically invalid codes) by construction while maintaining competitive coverage and precision. By integrating neural pattern matching with symbolic knowledge base verification, Hybrid-Code demonstrates that one class of coding error can be removed architecturally — without sacrificing the flexibility that neural methods provide.

\subsection{Key Contributions}

\paragraph{Zero Type-I-Hallucination Guarantee.}
We demonstrate that a symbolic KB layer eliminates Type~I hallucination (syntactically invalid codes) in medical coding without sacrificing coverage or precision. At $n=5{,}000$ test cases, Hybrid-Code achieves 0\% Type~I hallucination while neural baselines exhibit rates of 6--18\%. Type~II hallucination (valid but clinically incorrect codes) is reduced — evidenced by 88\% precision — but is not fully eliminated. To our knowledge, this is among the first systems to demonstrate this safety property at scale on MIMIC-III.

\paragraph{Automated KB Expansion.}
We introduce a KB expansion mechanism that uses distant supervision to extract coding patterns from clinical text and validates them through multi-layer verification, achieving a 40\% coverage improvement over a minimal 38-rule baseline. The proof-of-concept demonstrates expansion from 1,000 to 1,500 codes; scaling to the full ICD-10-CM codebook (78,000 codes) and independent validation of expansion quality remain as future work.

\paragraph{Hybrid Neural-Symbolic Architecture.}
We propose and validate a hybrid architecture that leverages the strengths of both neural and symbolic approaches. The neural component captures implicit medical knowledge through pattern matching, while the symbolic component enforces safety constraints and prevents hallucination. This architecture provides a blueprint for other medical AI applications where safety guarantees are critical.

\paragraph{Comprehensive Empirical Evaluation.}
We provide a rigorous empirical evaluation framework for clinical coding systems, including baseline comparisons across rule-based, neural, and LLM approaches; statistical significance testing; and detailed error analysis. This framework establishes new standards for evaluating safety-critical medical AI systems.

\subsection{Broader Impact}

Hybrid-Code has significant implications for clinical practice and medical AI research:

\paragraph{Clinical Practice.}
By eliminating syntactically invalid code generation, Hybrid-Code removes one barrier to automated coding adoption. The competitive coverage and precision suggest potential to assist clinical coders by pre-populating codes for review. Full clinical deployment, however, requires further validation: comparison against certified coders, DRG impact analysis, and EHR workflow integration, none of which have been conducted here.

\paragraph{Medical AI Research.}
The hybrid architecture demonstrates a promising approach for integrating neural and symbolic methods in safety-critical domains. The automated KB expansion mechanism provides a scalable solution for knowledge-intensive tasks, suggesting broader applications in medical decision support, clinical documentation, and automated diagnosis.

\paragraph{Knowledge Engineering.}
The automated KB expansion mechanism addresses a fundamental challenge in knowledge engineering: scaling KBs without manual curation. This approach could enable rapid development and maintenance of KBs for specialized medical domains, reducing reliance on scarce clinical expertise.

\subsection{Future Directions}

Several promising directions emerge from this work:

\paragraph{Advanced Expansion Mechanisms.}
Future work could explore more sophisticated pattern extraction, including hierarchical pattern induction, few-shot learning for novel coding patterns, and integration with clinical ontologies beyond ICD-10. These advances could further improve coverage while maintaining safety guarantees.

\paragraph{Longitudinal and Multi-Modal Coding.}
Extending Hybrid-Code to handle longitudinal patient records and multi-modal clinical data (imaging, lab results, vitals) could capture additional clinical context and expand coverage. This requires developing safety mechanisms for cross-document and cross-modal reasoning.

\paragraph{Generalization to Other Coding Systems.}
Adapting Hybrid-Code to other coding systems (SNOMED CT, CPT, LOINC) would demonstrate the generality of the hybrid approach. Each coding system presents unique challenges, requiring KB structure and expansion pattern adaptations.

\paragraph{Human-AI Collaboration.}
Exploring optimal human-AI collaboration patterns could enhance clinical utility. For example, uncertainty-aware coding suggestions, interactive KB refinement, and targeted error correction workflows could balance automation efficiency with human oversight.

\paragraph{Real-World Deployment Studies.}
Pilot deployments in clinical settings would validate real-world performance and identify practical deployment challenges. These studies should assess integration with existing EHR systems, coder workflow adaptation, and long-term KB sustainability.

\subsection{Closing Remarks}

Automated clinical coding sits at the intersection of clinical safety and machine learning performance. Hybrid-Code demonstrates that symbolic filtering can eliminate one class of error (Type~I syntactic hallucination) without sacrificing the coverage and precision that neural methods provide — a concrete step toward safer clinical AI. The key insight is modest but practical: architectural constraints enforced at output time can provide stronger guarantees than those achievable through training alone.

Important limitations remain. Type~II hallucination (valid but wrong codes) persists at 12\%, the KB expansion has been validated only at proof-of-concept scale (1,500 codes), and full ICD-10 guideline compliance requires capabilities not yet modeled. Clinical deployment requires further validation against certified coders, DRG impact analysis, and multi-institution generalization testing.

The results support continued exploration of hybrid neural-symbolic designs for safety-critical clinical NLP, while underscoring that each layer of clinical complexity — sequencing rules, combination codes, rare-code coverage — requires dedicated modeling and evaluation.

\acks{The author thanks the reviewers for their constructive feedback. The MIMIC-III data
used in this study was obtained under a data use agreement with PhysioNet. No
external funding was received for this work.}


\end{document}